\documentclass[aps,prb,twocolumn,showpacs,superscriptaddress]{revtex4}

\usepackage{graphicx} 
\begin{document}

\title{Monolayer honeycomb structures of group IV elements and III-V binary compounds}

\author{H. \c{S}ahin}
\affiliation{UNAM-Institute of Materials Science and
Nanotechnology, Bilkent University, 06800 Ankara, Turkey}

\author{S. Cahangirov}
\affiliation{UNAM-Institute of Materials Science and
Nanotechnology, Bilkent University, 06800 Ankara, Turkey}

\author{M. Topsakal}
\affiliation{UNAM-Institute of Materials Science and
Nanotechnology, Bilkent University, 06800 Ankara, Turkey}

\author{E. Bekaroglu}
\affiliation{UNAM-Institute of Materials Science and
Nanotechnology, Bilkent University, 06800 Ankara, Turkey}

\author{E. Akturk}
\affiliation{UNAM-Institute of Materials Science and
Nanotechnology, Bilkent University, 06800 Ankara, Turkey}

\author{R. T. Senger}
 \affiliation{Department of Physics, Izmir Institute of Technology,
35430 Izmir, Turkey}

\author{S. Ciraci}\email{ciraci@fen.bilkent.edu.tr}
\affiliation{UNAM-Institute of Materials Science and
Nanotechnology, Bilkent University, 06800 Ankara, Turkey}
\affiliation{Department of Physics, Bilkent University, 06800 Ankara, Turkey}

\date{\today}

\begin{abstract}
Using first-principles plane wave calculations, we investigate two
dimensional honeycomb structure of Group IV elements and their
binary compounds, as well as the compounds of Group III-V
elements. Based on structure optimization and phonon mode
calculations, we determine that 22 different honeycomb materials
are stable and correspond to local minima on the Born-Oppenheimer 
surface. We also find that all the binary compounds containing 
one of the first row elements, B, C or N have planar stable 
structures. On the other hand, in the honeycomb structures of Si, 
Ge and other binary compounds the alternating atoms of hexagons 
are buckled, since the stability is maintained by puckering. For 
those honeycomb materials which were found stable, we calculated 
optimized structures, cohesive energies, phonon modes, electronic 
band structures, effective cation and anion charges, and some elastic constants. The band gaps 
calculated within Density Functional Theory using Local Density 
Approximation are corrected by GW$_{0}$ method. Si and Ge in 
honeycomb structure are semimetal and have linear band
crossing at the Fermi level which attributes massless Fermion  
character to charge carriers as in graphene. However, all binary 
compounds are found to be semiconductor with band gaps
depending on the constituent atoms. We present a method to reveal 
elastic constants of 2D honeycomb structures from the strain 
energy and calculate the Poisson's ratio as well as in-plane 
stiffness values. Preliminary results show that the nearly 
lattice matched heterostructures of these compounds can offer new 
alternatives for nanoscale electronic devices. Similar to those of
the three-dimensional Group IV and Group III-V compound 
semiconductors, one deduces interesting correlations among the 
calculated properties of present honeycomb structures.
\end{abstract}

\pacs{73.22.-f, 61.48.De, 63.22.-m, 62.23.Kn}

\maketitle

\section{introduction}

Last two decades, nanoscience and emerging nanotechnologies have
been dominated by honeycomb structured carbon based materials in
different dimensionality, such as fullerenes, single and multi
walled carbon nanotubes, graphene and its ribbons. In particular,
graphene, a two dimensional (2D) honeycomb structure of carbon,
has been an active field of research.\cite{novo} Because of unique
symmetry, electron and hole bands of graphene show linear band
crossing at the Fermi level\cite{wallace} resulting in a massless
Dirac Fermion like behavior of charge carriers. As a result, Klein
paradox, an interesting result of quantum electrodynamics was expected
to be observed in graphene.\cite{kats,novo2, itzykson and
zuber,calogeracos} Moreover, it was shown that half-integer
quantization of Hall conductance\cite{novo2,zhang, berger} can be
observed in graphene. Unusual electronic and magnetic properties
of graphene, such as high carrier mobility, ambipolar effect, have
promised variety of applications. In addition to some early works
on crystalline order in planar structures,\cite{mermin,doussal}
possibility of very large one-atom-thick two dimensional (2D)
crystals with intrinsic ripples is reported
theoretically\cite{kastnelson-nature} and
experimentally.\cite{meyer} Not only extended 2D graphene sheets
but also quasi-one-dimensional graphene ribbons with armchair or
zigzag edges have shown unusual
electronic,\cite{son,barone,areshkin,sasaki,abanin,ezawa}
magnetic\cite{kobayashi,fujita,nakada1,wakabayashi1} and quantum
transport properties.\cite{graphene_applications2,
graphene_applications4,jie,hasan}

All these experimental and theoretical studies on graphene
created significant interest in one-atom-thick honeycomb lattices
of other Group IV elements and compounds of III-V and II-VI Group
elements. Recently, the boron-nitride (BN) honeycomb sheet
was reported as a stable ionic monolayer.\cite{bn-synthesis}
BN has the same planar structure as graphene with a nearest neighbor distance of 1.45 \AA. However, its ionic character causes a gap opening at the K-point. Thus, instead of being a semimetal, BN honeycomb structure is a wide band gap
insulator with an energy gap of 4.64 eV. Soon after its synthesis,
several studies on nanosheets \cite{bn-nanosheets1} and
nanoribbons\cite{Guo,louie-nanoletter,barone2,bn-mehmet} of BN
have been reported.

Hexagonal monolayer of Zinc Oxide (ZnO) is a II-VI metal-oxide analogue
of graphene and BN. Previously, works on nanostructures of ZnO
such as nanosheets,\cite{znosheet} nanobelts,\cite{znobelt}
nanotubes,\cite{znotube} nanowires\cite{znowire} and 
nanoribbons\cite{znomehmet} were reported and recently the
synthesis of ZnO bilayer honeycomb structure was also 
achieved.\cite{znobilayer} In contrast to graphene, ZnO
nanoribbons have ferromagnetic order in their ground state due to
electronic states at the zigzag edges dominated by oxygen 
atoms.\cite{mendez1,mendez2} 

Two dimensional SiC honeycomb sheet is another Group IV binary 
compound displaying interesting properties. While the infinite 
periodic 2D form of SiC is a semiconductor with 2.55 eV band gap, 
and its zigzag nanoribbons are magnetic metals, the armchair 
ribbons are nonmagnetic semiconductors.\cite{sic}  
Half-metallicity is also predicted for narrow SiC zigzag 
nanoribbons without any chemical decoration or applied external 
field. Furthermore, functionalization of SiC single sheets upon formation of various types of vacancies and adatom decoration was also predicted.\cite{erman-sic}

\begin{figure}
\includegraphics[width=8.5cm]{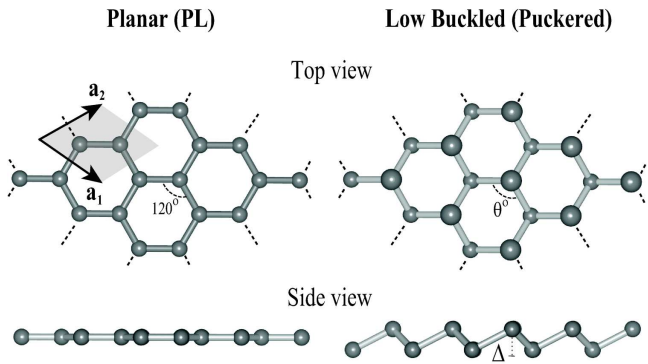}
\caption{ Top and side views for two dimensional (a) planar (PL) 
and (b) low-buckled (LB) (or puckered) honeycomb structure. In 
the PL structure atoms are located on the same plane. In the LB 
structure the alternating atoms are located in two different 
parallel planes. The buckling $\Delta$ is the distances between 
these two planes. Bravais lattice vectors for both structure are 
given with $|\vec{a_{1}}|=|\vec{a_{2}}|=a$. The unitcell is delineated and shaded.} \label{fig1}
\end{figure}

Very recently, we have reported that  among Group IV
elements, not only C but also Si and Ge can form stable honeycomb
structures.\cite{Si} It is found that for Si and Ge planar
(PL) geometry is not the lowest energy configuration and it
is not stable. Alternatively, it was shown that a low-buckled (LB)
(or puckered) geometry corresponds to a stable local minimum on
the Born-Oppenheimer surface. Buckled honeycomb structure of Si
was pointed out even in some earlier
studies.\cite{takeda,durgun,zhang} Surprisingly, in spite of their
puckered geometrical structure, Si and Ge monolayers have
electronic band structures which are similar to graphene. As a
result, linear crossing of $\pi$- and $\pi^{*}$-bands at K- and
K'-points of the hexagonal BZ attributes a massless Dirac fermion
character to the charge carriers. Quasi 1D honeycomb structures,
namely nanoribbons of Si and Ge, also show interesting electronic
and magnetic properties depending on their width and orientation.
Successful realization of single crystal silicon monolayer
structures\cite{nakamura,krishnan} through chemical exfoliation
shows that 2D silicon monolayers with their low resistivity and
extremely thin structures can be quite promising for
nanoelectronics.

\begin{table*}
\caption{Calculated results for Group IV elements, their binary
compounds and Group III-V compounds having honeycomb structure.
Stable structures are identified as PL or LB standing for the
planar and low-buckled geometries, respectively. The values of
angle between neighboring bonds, $ \theta$; buckling parameter, $\Delta$;
nearest neighbor distance, $d$; 2D hexagonal lattice constant,
$|\vec{a_{1}}|=|\vec{a_{2}}|=a$; cohesive energy, $E_{c}$; minimum
value of the energy gap, $E_{G}$ calculated using LDA and
corrected by GW$_{0}$ with the symmetry points indicating where
minimum (maximum) of conduction (valence) band occurs; calculated
effective charges on the constituent cation/anion, $Z^{*}_{c/a}$;
Poisson's ratio, $\nu$; and in-plane stiffness, $C$, are given. 
Some of the structural parameters are described in 
Fig.~\ref{fig1}.}

\label{table} \begin{tabular}{lccccccccccccc} \hline \hline
 &  &  &  &  &  &  &  &  &  &  &  &  & \tabularnewline
 & $ $ $Geometry$ $ $  & $ $ $ $ $\theta(deg)$ $ $ $ $  & $ $ $\Delta(\AA)$ $ $  & $ $ $d(\AA)$ $ $  & $ $ $a(\AA)$ $ $  & $ $ $E_{c}$ (eV) $ $ & $E_{G}(eV)$  & $Z^{*}_{c/a}$ &  & $\nu$ &  & C (J/m$^{2}$) & \tabularnewline
 &  &  &  &  &  &  & LDA $ $ $ $ $ $ $ $ $GW_{0}$  &  &  &  &  &  & \tabularnewline
\hline
\hline
 &  &  &  & $GROUP$ & $IV$ &  &  &  &  &  &  &  & \tabularnewline
\hline
\hline
\textbf{Graphene}  & PL  & 120.0  & -  & 1.42  & 2.46  & 20.08  & semimetal  & 0.0/0.0  &                            & 0.16 &  & 335 & \tabularnewline
\hline
\textbf{Si}        & LB  & 116.4  & 0.44 & 2.25  & 3.83  & 10.32  & semimetal  & 0.0/0.0  &                          & 0.30 &  & 62 & \tabularnewline
\hline
\textbf{Ge}        & LB  & 113.0  & 0.64 & 2.38  & 3.97  & 8.30  & semimetal & 0.0/0.0  &                            & 0.33 &  & 48 & \tabularnewline
\hline
\textbf{SiC}       & PL  & 120.0  & -  & 1.77  & 3.07  & 15.25  & 2.52/KM-3.89/KM$ $ $ $ $ $ $ $& 1.53/6.47$ $ $ $ &                     & 0.29 &  & 166 & \tabularnewline
\hline
\textbf{GeC}       & PL  & 120.0 & -  & 1.86  & 3.22  & 13.23 & 2.09/KK-3.56/KK & 2.82/5.18  &                       & 0.33 &  & 142 & \tabularnewline
\hline
\textbf{SnGe}      & LB  & 112.3 & 0.73  & 2.57  & 4.27  & 8.30  & 0.23/KK-0.41/KK & 3.80/4.20  &                    & 0.38 &  & 35 & \tabularnewline
\hline
\textbf{SiGe}      & LB  & 114.5 & 0.55  & 2.31  & 3.89  & 9.62  & 0.02/KK-0.00/KK & 3.66/4.34  &                    & 0.32 &  & 57 & \tabularnewline
\hline
\textbf{SnSi}      & LB  & 113.3 & 0.67  & 2.52  & 4.21  & 8.72  & 0.23/KK-0.69/KK  & 3.89/4.11  &                   & 0.37 &  & 40 & \tabularnewline
\hline
\textbf{SnC}       & PL  & 120.0 & -  & 2.05  & 3.55  & 11.63 & 1.18/$\Gamma$K-6.01/$\Gamma$K & 2.85/5.15  &         & 0.41 &  & 98 & \tabularnewline
\hline
\hline
 &  &  &  & $III-V$ &  $GROUP$&  &  &  &  &  &  &  & \tabularnewline
\hline
\hline
\textbf{BN}  & PL  & 120.0  & -  & 1.45  & 2.51  & 17.65 & 4.61/KK-6.57/$\Gamma$K & 0.85/7.15  &                    & 0.21 &  & 267 & \tabularnewline
\hline
\textbf{AlN}  & PL  & 120.0  & -  & 1.79  & 3.09  & 14.30 & 3.08/$\Gamma$M-5.74/$\Gamma$M & 0.73/7.27  &            & 0.46 &  & 116 & \tabularnewline
\hline
\textbf{GaN}  & PL  & 120.0  & -  & 1.85  & 3.20  & 12.74  & 2.27/$\Gamma$K-4.58/$\Gamma$K  & 1.70/6.30  &          & 0.48 &  & 110 & \tabularnewline
\hline
\textbf{InN}  & PL  & 120.0  & -  & 2.06  & 3.57  & 10.93  & 0.62/$\Gamma$K-5.35/$\Gamma$$\Gamma$  & 1.80/6.20  &   & 0.59 &  & 67 & \tabularnewline
\hline
\textbf{InP}  & LB  & 115.8  & 0.51  & 2.46  & 4.17  & 8.37  & 1.18/$\Gamma$K-2.68/$\Gamma$K  & 2.36/5.64  &        & 0.43 &  & 39 & \tabularnewline
\hline
\textbf{InAs}  & LB  & 114.1  & 0.62  & 2.55  & 4.28  & 7.85  & 0.86/$\Gamma$$\Gamma$-1.92/$\Gamma$$\Gamma$  & 2.47/5.53  &  & 0.43 &  & 33 & \tabularnewline
\hline
\textbf{InSb}  & LB  & 113.2  & 0.73 & 2.74  & 4.57  & 7.11  & 0.68/$\Gamma$$\Gamma$-1.71/$\Gamma$$\Gamma$  & 2.70/5.30  &  & 0.43 &  & 27 & \tabularnewline
\hline
\textbf{GaAs}  & LB  & 114.7  & 0.55  & 2.36  & 3.97  & 8.48  & 1.29/$\Gamma$K-2.77/$\Gamma$K  & 2.47/5.53  &       & 0.35 &  & 48 & \tabularnewline
\hline
\textbf{BP}  & PL  & 120.0  & -  & 1.83  & 3.18  & 13.26  & 0.82/KK-1.67/KK  & 2.49/5.51  &                         & 0.28 &  & 135 & \tabularnewline
\hline
\textbf{BAs}  & PL  & 120.0  & -  & 1.93  & 3.35  & 11.02 & 0.71/KK-1.14/KK  & 2.82/5.18  &                         & 0.29 &  & 119 & \tabularnewline
\hline
\textbf{GaP}  & LB  & 116.6  & 0.40 & 2.25  & 3.84  & 8.49  & 1.92/$\Gamma$K-3.50/KM  & 2.32/5.68  &                & 0.35 &  & 59 & \tabularnewline
\hline
\textbf{AlSb}  & LB  & 114.8  & 0.60  & 2.57 & 4.33  & 8.04  & 1.49/KM-2.01/KK  & 1.58/6.42  &                      & 0.37 &  & 35 & \tabularnewline
\hline
\textbf{BSb}  & PL  & 120.0  & -  & 2.12  & 3.68  & 10.27  & 0.39/KK-0.23/KK  & 3.39/4.61  &                        & 0.34 &  & 91 & \tabularnewline
\hline\hline &  &  &  &  &  &  &  &  &  &  &  &  & \tabularnewline
\end{tabular}
\end{table*}

Motivated by the recent experimental developments and theoretical
investigations on 2D monolayer honeycomb structures, in
this paper we carried out a systematic study of similar
structures of Group IV elements and III-V binary compounds based
on first principles calculations within Density Functional Theory
(DFT). Our objective is to reveal whether monolayer honeycomb structures can be found as a local minimum on the Born-Oppenheimer surface. We
hope that the predictions of our work will guide further
experimental studies towards the synthesis of new materials having
honeycomb structure. The present work, which considers a total of 26 elemental and binary compounds in 2D honeycomb structure and
reveals whether they are stable, is an extension to our
preliminary work on Si and Ge puckered honeycomb structures.
\cite{Si} Based on extensive analysis of stability, 22
different materials out of 26 are found to be stable in a local
minimum on the Born-Oppenheimer surface either in finite size or
in infinite periodic form. We hope that interesting properties
predicted by this study will promote efforts towards synthesizing
new materials and heterostructures, which will constitute a
one-dimensional analogue of 3D family of tetrahedrally coordinated
semiconductors.

The organization of this paper is as follows; In Sec. II the
methods together with parameters used in our calculations are
outlined. In Sec. III, we determine the atomic structure and
related lattice constants of the honeycomb structures via total
energy minimization. We also discuss how the stability of the
structure is maintained through puckering. In the same section, we
present our results regarding the calculation of phonon modes
and our analysis of stability based on these results. The
mechanical properties of these structures are investigated in Sec.
IV. We discuss the electronic band structure of
various stable materials calculated within DFT in Sec. V. The underestimated
band gaps are corrected by using GW$_{0}$ calculations. As a proof
of concept for a possible future application of these materials we
consider semiconductor superlattices formed from the periodically
repeating pseudomorphic heterostructure in Sec VI. We showed that the superlattices have an electronic structure different from those of constituent materials and behave as multiple quantum
well structures with confined states. In Sec. VII, our
conclusions are presented.

\section{Methods}

We have performed self-consistent field, first-principles plane-wave
calculations\cite{vasp1,vasp2} within DFT for total energy and
electronic structure calculations. Projector augmented wave (PAW)
potentials\cite{blochl} and exchange-correlation potentials
approximated by local density approximation\cite{lda} (LDA) are
used in our calculations. In the self-consistent field potential
and total energy calculations a set of (25x25x1) \textbf{k}-point
sampling is used for Brillouin Zone (BZ) integration in
\textbf{k}-space. Kinetic energy cutoff $ \hbar^2
|\mathbf{k}+\mathbf{G}|^2 / 2m $ for plane-wave basis set is taken
as 500 eV . The convergence criterion of self consistent
calculations is $10^{-5}$ eV for total energy values. By using the
conjugate gradient method, all atomic positions and unitcell were
optimized until the atomic forces were less than 0.05 eV/\AA.
Pressures on the lattice unit cell are decreased to values less
than 0.5 kB. To prevent interactions between the adjacent
supercells a minimum of 10 $\AA$ vacuum spacing is kept.

To correct the energy bands and band gap values obtained by LDA,
frequency-dependent GW$_{0}$ calculations are carried
out.\cite{gw} Screened Coulomb potential, W, is kept fixed to
initial DFT value W$_{0}$ and Green's function, G, is iterated
five times. Various tests are performed regarding vacuum level,
kinetic energy cut-off potential, number of bands,
\textbf{k}-points and grid points. Final results of
GW$_{0}$ corrections are obtained by using (12x12x1) \textbf{k}-points in BZ, $15$~\AA~ vacuum spacing, default cut-off potential for GW$_{0}$, 160 bands and 64 grid points.

Cohesive energies  ($E_c$) per pair of atoms (see Table
\ref{table}) are calculated by using the expression

\begin{equation}\label{equ:binding}
E_c={E_{T}[AB]} - E_{T}[A] - E_{T}[B]
\end{equation}

where $E_{T}[AB]$ is the total energy per A-B pair of the
optimized honeycomb structure; $E_{T}[A]$ and $E_{T}[B]$ are the
total energies of free A and B atoms. All of them are calculated 
in the same cell. For graphene, Si and Ge, $A=B$. For the charge 
transfer analysis, the effective charge on atoms are obtained by 
Bader method.\cite{bader} In fact, various methods 
for charge transfer analysis give similar trends for the 
honeycomb structures studied in this paper. For rigorous test of 
the stability of  fully relaxed honeycomb structures under study, 
we also calculated phonon modes by using force constant 
method.\cite{alfe} Here the Dynamical Matrix was constructed from forces, 
resulting from displacements of certain atoms in (7x7) supercell, 
calculated by VASP software.

\section{Atomic Structure}

We first present a detailed analysis of two-dimensional (2D)
hexagonal structure of binary compounds of Group IV elements,
their binary compounds and Group III-V compounds all forming
honeycomb structure. In our study, we also include the discussion
of graphene, Si and Ge in honeycomb structure for the sake of
comparison.

\begin{figure}
\includegraphics[width=8.5cm]{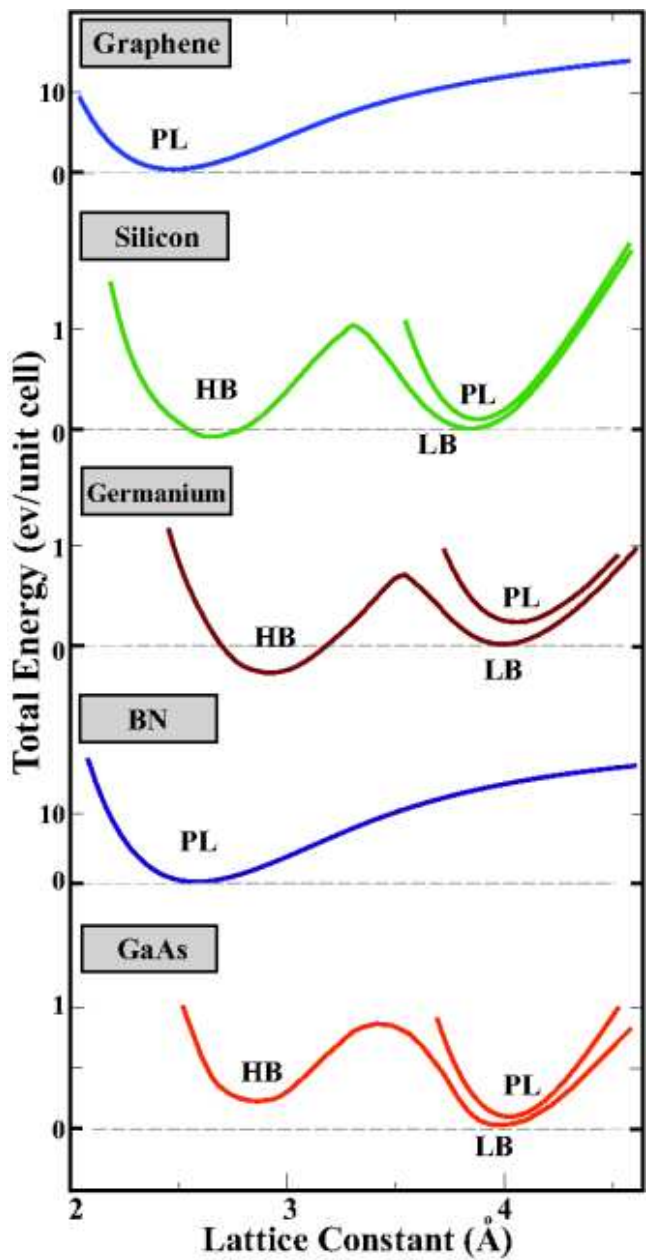}
\caption{(Color online) Variation of total energy of C, Si, Ge,
BN, GaAs honeycomb structures with respect to the lattice constant
$a$ of 2D hexagonal lattice. The stable local minima of the Born-Oppenheimer surface for each structure is shown with a dashed line separately.}
\label{fig2}
\end{figure}

Graphene has a 2D hexagonal lattice in which C atoms are arranged
to form a planar (PL) honeycomb structure as shown in
Fig.\ref{fig1}. Accordingly, it has a six fold rotation axis, 
$C_{6}$ at the center of the hexagon, which is perpendicular to 
the atomic plane. Hexagonal lattice has a two-atom basis in the 
primitive unit cell, corresponding to A- and B-sublattices. That 
is three alternating atoms of each hexagon belong to one of the 
two sublattices. In graphene planar geometry is assured by the 
formation of strong $\pi$-bonding between two nearest neighbor 
$p_{z}$-orbitals perpendicular to the graphene plane. The 
resulting $\pi$- and $\pi^{*}$-bands determine also relevant 
electronic properties. In addition, there are strong yet 
flexible, covalent $\sigma$-bonds derived from the planar hybrid 
$sp^{2}$ orbitals between adjacent C atoms. Nearest C atoms are 
separated by 1.42 \AA~and the magnitude of the hexagonal Bravais 
lattice vector is 2.46~\AA. Briefly, the planar $sp^{2}$ 
hybridization and perpendicular $p_{z}$ orbitals underlie planar 
geometry, unusual mechanical strength and electronic structure of 
graphene.

In Fig.~\ref{fig2} we show the variation of the total energy with
respect to the lattice constant $a$ of the 2D hexagonal Bravais
lattice. We see that C and BN stayed planar and have a single 
minimum. The situation with Si, Ge and GaAs is different, since they have two other minima corresponding to low buckled (LB) and high buckled (HB) geometries in addition to planar geometry. In fact, the total energies corresponding to the minimum of the planar geometry are already higher than those of LB and HB geometries. In buckled geometries, while atoms of A-sublattice are rising, those of B-sublattice are lowered. At the end atoms of A- and B-sublattices lie in different planes having a buckling distance, $\Delta$ as shown in Fig.~\ref{fig1}. The value of $\Delta$ in HB geometry is high and is in the ranges of $\sim$ 2.5\AA, but it is low in LB geometry and ranges between 0.4 and 0.7\AA. We note that two minima  corresponding to HB and LB geometries in Fig.~\ref{fig2} are seperated by a significant energy barrier. As we discuss in subsection III.B the minimum of HB is not actually a local minimum on Born-Oppenheimer surface. The six-fold rotation symmetry of 
graphene is broken as a result of buckling and changes to the 
three-fold rotation symmetry $C_{3}$. The similar symmetry 
breaking takes place also in Group III-V compounds having PL 
honeycomb structures.

In concluding this discussion, we point out that even
if the calculated total energy has a minimum relative to a
specific structural parameter, this may not correspond to a local
minimum. Then it remains to answer which of these minima in
Fig.~\ref{fig2} corresponds to a local minima on the
Born-Oppenheimer surface. At this point, reliable tests for 
stability of structure have to be performed.

\subsection{Phonon modes and stability}

\begin{figure*}
\begin{center}
\includegraphics[width=18cm]{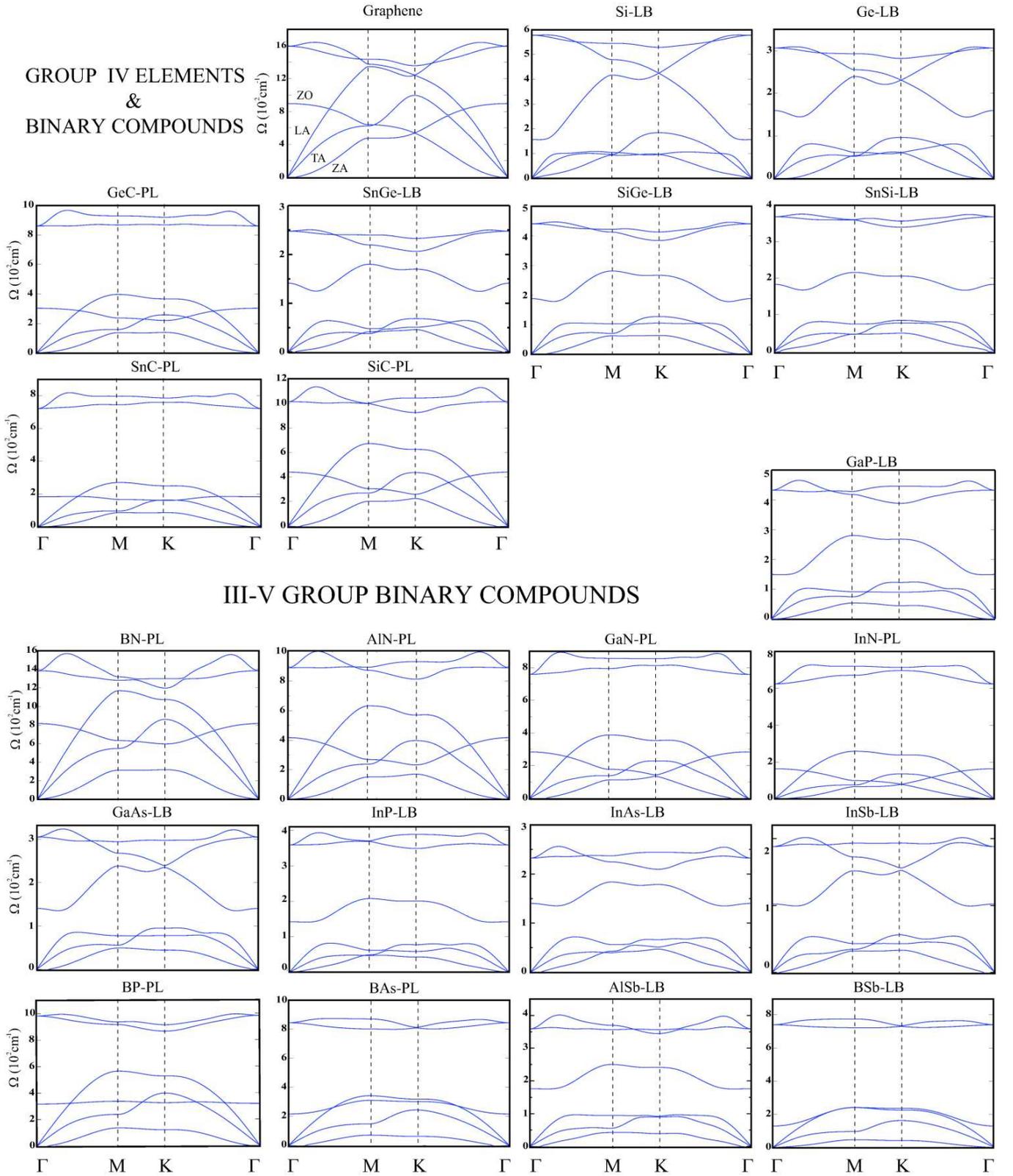}
\caption{ Calculated vibration frequencies of phonon modes
$\Omega$ versus \textbf{k} of Group IV elements, their binary
compounds and Group III-V compounds having honeycomb structure.
Compounds having at least one constituent from first row elements
have tendency to form planar structure. PL and LB stand for the
planar and low-buckled (puckered) honeycomb structures,
respectively.} \label{fig3}
\end{center}
\end{figure*}

Analysis of phonon modes provides a reliable test for a
structure optimized conjugate gradient method. If there is an
instability related with a phonon mode with \textbf{k} in BZ, the
square of frequency, $\Omega$(\textbf{k}) obtained from the
dynamical matrix becomes negative implying an imaginary frequency.
Phonon calculations are performed by taking into account the
interactions in (7x7x1) large supercells consisting of 98 atoms.
For all the infinite 2D honeycomb structures there are three
acoustical (A) and three optical (O) modes.
In Fig. \ref{fig3}, we present the calculated dispersions of phonon
modes of various honeycomb structures; namely Group IV elements
and their binary compounds, as well as Group III-V compounds.
According to these results some materials, for example Sn cannot
be stable in the honeycomb structure. None of the honeycomb structures are stable in the HB geometry. Only 22 honeycomb structures studied in
this paper display $\Omega$(\textbf{k}) dispersions, which confirm
their stability either in PL or in LB geometry.

Well known linear behavior of phonon dispersions of LA and TA
branches and quadratic behavior of ZA branch around $\Gamma$-point
also exist for all 2D lattices. Among these, the LA and TA phonon
branches are heat carrying modes. However, it was shown that
bending branch ZA makes negligible contribution to thermal
conductivity.\cite{klemens} Note that only the ZA branch gets
imaginary frequencies for \textbf{k} wave vectors in the vicinity
of $\Gamma$-point in BZ. While this indicates the structural
instability for infinitely large periodic structures, it can be
taken also as evidence that an infinitely large 2D hexagonal
lattice can be stabilized by defects or ripples with large wave
length. Our analysis for such systems treated in ($n \times n$)
supercells showed the existence of ripples. Clearly, finite
patches of such structures can become stable. This argument has
been confirmed by our calculation of vibration modes of finite
size patches.

In our earlier paper\cite{Si} we found that ZA mode of 2D
periodic Ge honeycomb gets imaginary frequencies near
$\Gamma$-point of BZ. This situation has been interpreted as the
instability against long wavelength transversal waves, which can
be stabilized by rippling or by limiting the size of Ge sheets. An extensive analysis of phonon modes in the present study revealed that the extent of the region of imaginary frequencies around the $\Gamma$-point also depends on the mesh size used in the calculations. Decreasing the mesh size may lead to the decreasing of their particular zone. Therefore, a tedious analysis of the right mesh size is required to determine whether or not the imaginary frequency zone of ZA mode is an artifact of the mesh size. We performed this analysis of mesh size for Ge-LB structure and found an optimum fine mesh size where imaginary frequencies of ZA mode disappeared. We also note that since the interatomic forces related with ZA modes decay rapidly, the numerical inaccuracy in calculating forces due to the transversal displacement of distant atoms may give rise to difficulties in the treatment of ZA modes. Briefly, caution has to be exercised in deciding whether the imaginary frequencies of ZA modes is an artifact of numerical calculations. Of course, if the presence of imaginary frequencies near the $\Gamma$-point is a reality, the instability of infinite and 2D periodic structures can be stabilized by finite size or large wavelength ripplings.

Calculated phonon dispersion of graphene is in good agreement with
previous LDA results and also with reported experimental
data.\cite{ihlee,narasimhan,yan} Around 1600 $cm^{-1}$, LO and TO
eigenmodes are degenerate at $\Gamma$ point. In-plane TA and LA
eigenmodes have linear dispersion around the $\Gamma$ point. As it
is mentioned in earlier works on 2D structures, out-of-plane ZA
eigenmode have quadratic phonon dispersion in the vicinity of
$\Gamma$-point. Here, the calculated value for
out-of-plane optical eigenmode ZO is around 900 $cm^{-1}$.
Existence of strong electron-phonon coupling in TO eigenmode at
the $K$-point and $E_{2g}$ modes at $\Gamma$ -point is the reason
of the Kohn anomaly at these points. Therefore, scattering by
phonon with the energies that corresponds to these modes can cause
noticeable decrease in transmission spectrum.\cite{yan,pisanec}
Since force constant decreases with increasing atomic number or
row number in the Periodic Table, calculated vibration frequencies
exhibit the same trend.

As a result of symmetry in honeycomb structures of Group
IV elements (such as graphene, Si and Ge), ZO and TO
branches cross at $K$ point. We also note that the ZO branch of a
binary compound comprising at least one element from the first row
falls in the frequency range of acoustical vibration modes. By
comparing the phonon dispersions of InN, InP, InAs and InSb
samples, it is seen that ZO mode have increasing tendency to move
apart from the LA and TA modes with increasing nearest neighbor
distance. However, in all the samples containing first row
elements, ZO mode is located between the LA and TA modes. These
characteristic trends of ZO mode exists for all the 2D honeycomb
structures.

\subsection{Stability via Puckering}

\begin{figure}
\includegraphics[width=8.5cm]{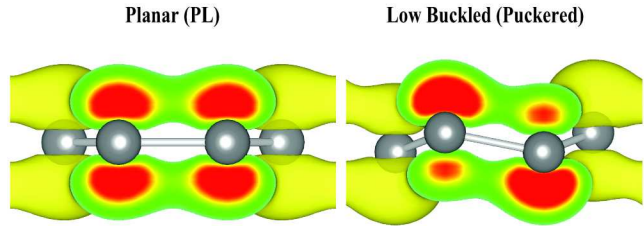}
\caption{In the honeycomb structures of C, Bn and several others, the planar geometry
is maintained by the strong $\pi$-bonding through the perpendicular
$p_{z}$ orbitals, in addition to the $\sigma$ bonding through
the $sp^{2}$ hybrid orbitals. In the case of honeycomb structures
formed by the elements beyond the first row, the $\pi$-$\pi$
bonding is weakened due to the increasing bond length. The
structure is stabilized by puckering where $sp^{2}$ hybrid
orbitals are slightly dehybridized to form $sp^{3}$-like orbitals.
This situation is depicted for Si honeycomb structure.
Increasing charge density is plotted with colors from yellow
(light) to red (dark).  }
\label{fig4}
\end{figure}

According to analysis of stability based on the calculated phonon
modes, structures which do not contain first row elements occur 
in LB (puckered) structure corresponding to a local minimum in
Born-Oppenheimer surface. Through puckering the character of the
bonding changes. Different hybrid orbitals underlie the different
allotrophic forms of C atom. While the bonding of diamond structure
is achieved by tetrahedrally coordinated, directional $sp^{3}$
hybrid orbitals, $sp^{2}+p_{z}$ and $sp+p_{x}+p_{y}$ hybrid
orbitals make the bonding in graphene and cumulene (monatomic
chain of carbon atoms), respectively. In forming hybrid orbitals
one of two valence $s$ states is excited to $p$ state, whereby a
promotion energy is implemented to the system. However, by $s$ and
$p$ hybridization the hybrid orbitals yield the maximum overlap
between adjacent C-C atoms and hence the strongest possible 
bonding. This way, the promotion energy is compensated and the system attains
cohesion. In $sp^{3}$ hybrid combination one $s$ orbital is
combined with $p_x$, $p_y$ and $p_z$ orbitals to form four
orbitals directed from the central C atom towards its four
nearest neighbors in tetrahedral directions. The angle between
these bonds is $\sim 109.5^{\circ}$. In $sp^{2}$ one $s$ is
hybridized with $p_{x}$ and $p_{y}$ orbitals to make three planar
$sp^{2}$ which are directed from the central C atom at the corners
of the hexagons to its three nearest neighbors. For the cumulene
$s$ orbital is hybridized with $p_{z}$ orbital along the chain
axis. In this respect, the strengths (i.e. self-energy) of these
hybrid orbitals decreases with increasing number of $p$-type
orbitals in the combination; namely $sp$ is strongest whereas
$sp^{3}$ is least strongest. As for the the distance of C-C bonds,
it is shortest in cumulene (1.29), but longest in diamond (1.53
\AA). In addition to these hybrid orbitals, dangling $p$ orbitals
make also $\pi$-bonding between two C atoms. The
$\pi$-bonding between two adjacent C atoms in graphene and
cumulene assures the planarity and linearity, respectively. In
Fig.\ref{fig4} the charge density of the $\pi$-bonds between
neighboring C atoms explains how the stable planar geometry is
maintained.

As the bond distance between two nearest neighbor atoms increases
the overlap of the $p_{z}$ orbitals decreases. This, in turn,
decreases the strength of the $\pi$-bond. This is the situation in
the honeycomb structures of Si, where the  Si-Si bond distance
(2.34 \AA) increased by 92\% relative to that of the C-C bond. As
a result of weaker $\pi$ bonds in Si can not maintain the stability of the planar geometry, and the structure attains the stability through puckering, where three alternating atoms of a
hexagon raises as the remaining three is lowered. At the end the
structure is buckled by $\Delta$. Through buckling the $sp^{2}$
hybrid orbital is dehybridized and $s$, $p_{x}$, $p_{y}$ orbitals
are then combined with $p_{z}$ to form $sp^{3}$ hybrid orbitals.
While three $sp^{3}$ hybrid orbitals form covalent bonds with
three nearest neighbor atoms, one $sp^{3}$ orbital directed
upwards perpendicular to the atomic plane form a weak bond with
the adjacent $sp^{3}$ orbital directed downwards. Eventually, a
puckered Si honeycomb structure is reminiscent of
graphane\cite{elias, sofo, hasan-graphane} with alternating C
atoms saturated with hydrogen atoms from different sites. Briefly,
puckering occurs as a result of weakening of $\pi$-bonds, whereby
the structure regains its stability through tetrahedrally
coordinated $sp^{3}$-like bonding. Puckering may be explained in terms of Jahn-Teller theorem\cite{jahn-teller} predicting that an
unequal population of degenerate orbitals in a molecule leads to a
geometric distortion. This way the degeneracy is removed and the
total energy is lowered.

As shown in Table \ref{table}, 11 out of 22 honeycomb structures
prefer planar geometry, the rest is puckered to regain stability.
There are interesting examples for planar and puckered ring
structures: Besides BN, planar $B(CH_{3})_{3}$ molecule and well
known $B(OH)_{3}$ boric acid planar crystals are other examples
for boron including materials. $S_{n}$ rings such as $S_{7}$,
$S_{8}$ and $S_{12}$ have also puckered structures with a crown
shape.\cite{bmeyer} Another example is planar cyclobutadiene
($C_{4}H_{4}$); the well known molecule was shown that it lowers its energy when it has puckered (butterfly) structure for its
positive dianion ($C_{4}H_{4}^{+2}$). Additionally, it was shown
before\cite{chem-bond-theory} that puckered shape of
cyclooctatetraene $C_{8}H_{8}$ takes planar shape for its negative
dianion $C_{8}H_{8}^{-2}$. For the phosphazenes ($NPX_{2}$), while
$N_{4}P_{4}F_{8}$ is planar $N_{4}P_{4}Cl_{8}$ and
$N_{4}P_{4}(CH_{3})_{8}$ have buckled shapes\cite{labarre,
hazenkampf} and thus the rule regarding the compounds including 
first row elements is still valid.

Having discussed the general aspects, we now concentrate on the 
optimized atomic structure and corresponding electronic 
properties of stable honeycomb structures. Calculated values of 
atomic and electronic structural parameters are given In Table \ref{table}. 
One notes that 11 structure having planar geometry has at least 
one constituent from the first row elements of the Periodic 
Table; namely C, B and N. Since the radii of these atoms are 
relatively small, their presence as one of the constituents 
assures that the bond length is small enough to keep strong 
$\pi$-bonding. This explains how the radius of constituent
atoms enters as a crucial ingredient in the structure. The rest 
of the honeycomb structures in Table \ref{table} including Si and 
Ge are puckered to have LB geometry.

Finally, we note that calculated results given in Table 
\ref{table} display interesting trends depending on the radius of 
constituent elements or their row number in the Periodic Table. 
For example, the bond strength or cohesive energy $E_{c}$ of a 
honeycomb structure gets weaker as the atomic radii or the row 
number of the constituent elements increase. Also band gap 
$E_{G}$ and lattice constant $a$ show similar trends.

\section{Mechanical Properties}

\begin{figure}
\includegraphics[width=8.5cm]{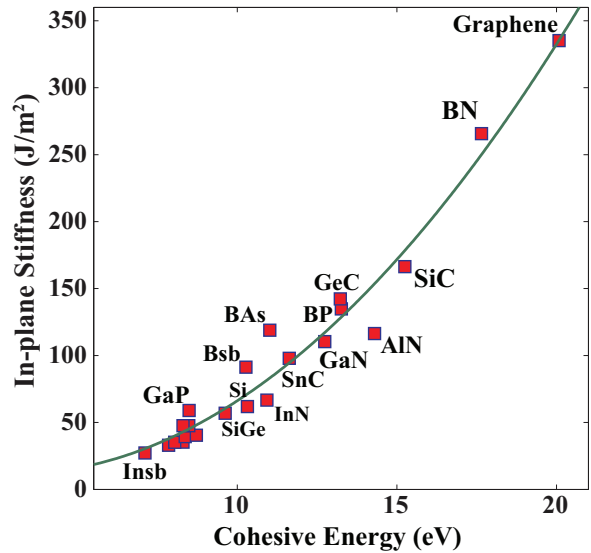}
\caption{(Color online) A plot showing the correlation between
the cohesive energy $E_c$ and in-plane stiffness $C$ among honeycomb
structures.}
\label{fig5}
\end{figure}

Honeycomb structure with $sp^{2}$ bonding underlies the unusual
mechanical properties providing very high in-plane strength, but
transversal flexibility. We note that graphene and
its rolled up forms, carbon nanotubes are among the strongest and
stiffest materials yet discovered in terms of tensile strength and
elastic modulus. We investigated the mechanical properties of 22 
stable honeycomb structures listed in Table \ref{table}. We 
focused on the harmonic range of the elastic deformation, where 
the structure responded to strain $\epsilon$ linearly. We pulled 
the rectangular unit cell in $x$- and $y$-directions in various 
amounts and generated a mesh of data corresponding to the strains 
in $x$ and $y$ directions versus strain energy defined as 
$E_{s}=E_{T}(\epsilon)-E_{T}(\epsilon=0)$;
namely, the total energy at a given strain $\epsilon$ minus the
total energy at zero strain. The data is fitted to a 
two-dimensional quadratic polynomial expressed by

\begin{equation}\label{equ:stiffness}
E_{S}(\epsilon_{x},\epsilon_{y})=a_{1}\epsilon_{x}^{2}+a_{2}\epsilon_{y}^{2}+a_{3}\epsilon_{x}\epsilon_{y}
\end{equation}

where $\epsilon_{x}$ and $\epsilon_{y}$ are the small strains
along $x$- and $y$-directions in the harmonic region. Owing to
the isotropy of the honeycomb structure $a_{1}=a_{2}$. The same
equation can be obtained from elastic tensor\cite{nye}
in terms of elastic stiffness constants, namely $a_{1} = a_{2} = 
(h\cdot A_{0}/2) \cdot C_{11}$ ; $a_{3}=(h \cdot A_{0}) \cdot C_{12}$.
Hence one obtains Poisson's ratio $\nu$ = -$\epsilon_{trans}$/$\epsilon_{axial}$, 
which is equal to $C_{12}/C_{11}=a_{3}/2a_{1}$. Similarly, the in-plane stiffness,  $C=h\cdot C_{11}\cdot(1-(C{}_{11}/C_{12})^{2}$) = (2a$_{1}$-(a$_{3}$)$^{2}$/2a$_{1}$)/(A$_{0}$).
Here $h$ and $A_{0}$ are the effective thickness and equilibrium area of the
system, respectively. In Table \ref{table}, calculated Poisson's ratio and in-plane
stiffness results are shown. The calculated value of the
in-plane stiffness of graphene is in agreement with the experimental value
of 340 (N/m).\cite{gr_exp} Graphene has highest in-plane
stiffness and lowest Poisson's ration among all honeycomb
structures of Group IV elements and Group III-V compounds. Being a
compound of first row elements, BN has second highest $C$ and
second lowest $\nu$. The Poisson's ratio $\nu$ increases with increasing row number of
elements of elemental and compound honeycomb structures. $C$ shows
a reverse trend. The order of values of $C$ in the last column of
Table \ref{table} is similar to that of cohesive energies $E_{c}$
in the seventh column. This clearly indicates a correlation
between $E_{c}$ and $C$ as shown in Fig.~\ref{fig5}.

\section{Electronic Structure}

\begin{figure*}
\begin{center}
\includegraphics[width=18cm]{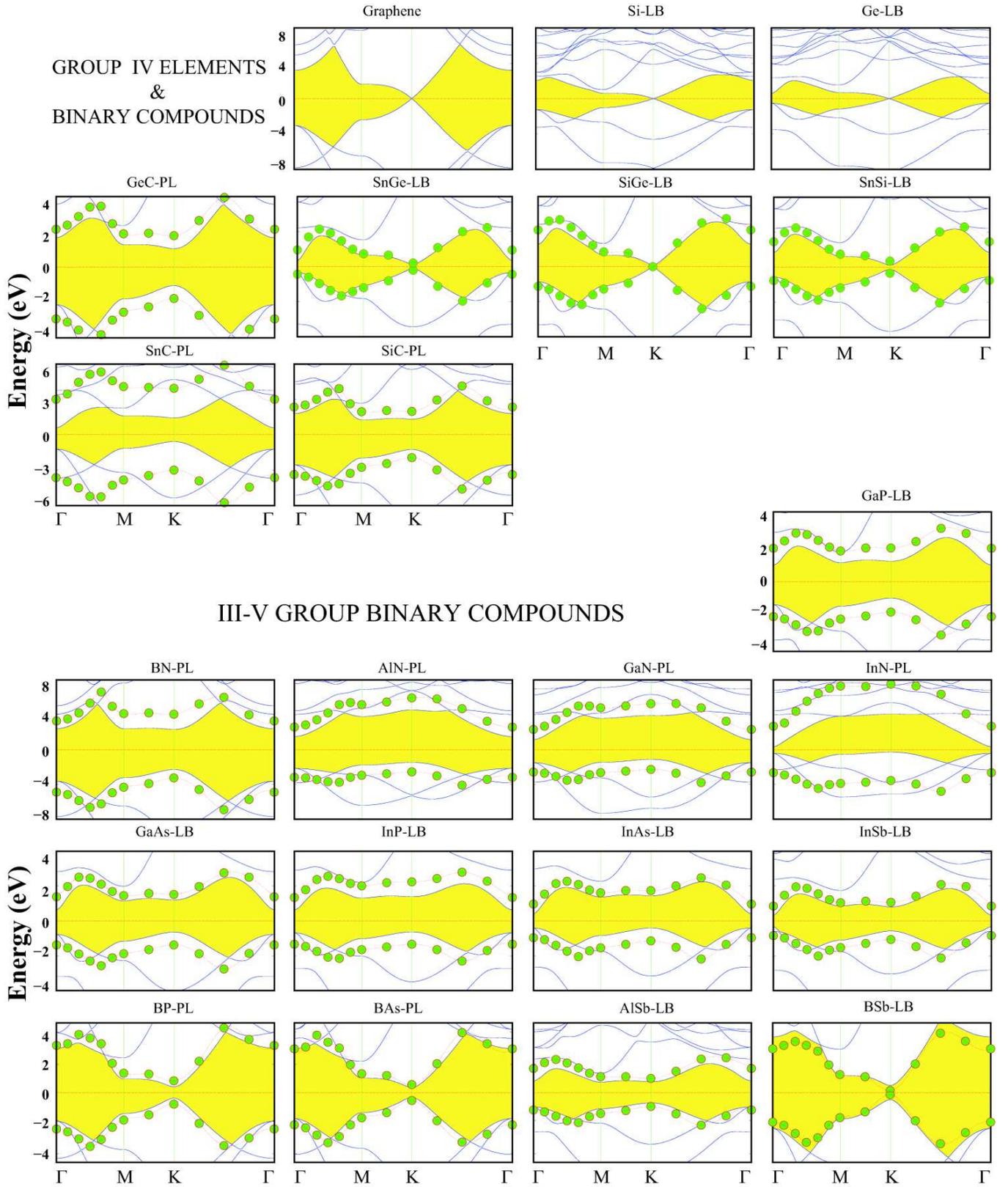}
\caption{(Color online) Energy bands of monolayer honeycomb
structures of Group IV elements and binary compounds between
different Group IV elements and Group III-V elements. All the
Group IV elements have semimetallic electronic structure. Band
structures show that like graphene $\pi$ and $\pi^{*}$ bands of
Si and Ge have linear band crossing at the Fermi level.
Binary compounds of Group IV and Group III-V elements are
semiconductors. Corrections LDA band gaps using GW$_{0}$ are
indicated by small circles. Band gaps are shaded.}
\label{fig6}
\end{center}
\end{figure*}

Our results on the electronic band structure of Group IV elements
and binary compounds between different Group IV elements and Group
III-V elements, which are stable in either infinite periodic form
or in finite size, are presented in Fig. \ref{fig6}. In these
hexagonal lattice structures (PL or LB) relevant electronic energy
bands around the Fermi level are derived from $\pi$ and
$\pi^{*}$-bands. In elemental honeycomb structures, such as 
graphene, Si and Ge, these bands have linear crossings at two 
in-equivalent $K$- and $K'$-points of BZ, called Dirac points and 
hence they are semimetallic. Because of their linear dispersion 
of $E$(\textbf{k}), the charge carriers near the Dirac points 
behave as massless Dirac Fermions. By fitting the $\pi$-and 
$\pi^{*}$- bands at $\bf{k}=\bf{K}+\bf{q}$ to the
expression,

\begin{equation}
E(\bf{q}) \simeq v_{F}\hbar|q|+O(q^{2})
\end{equation}

and neglecting the second order terms with respect to $q^{2}$,
one can estimate the Fermi velocity for both Si and Ge as
$v_{F} \sim 10^{6}$. We note that $v_{F}$ calculated for 2D
LB honeycomb structures of Si and Ge are rather high and close
to that calculated for graphene using the tight-binding bands.
It is also worth noting that because of the electron-hole
symmetry at K- and $K^{\prime}$-points of BZ, 2D LB Si
and Ge are ambipolar for E{(\bf{q})}= $E_{F} \pm \delta E$,
$\delta E$ being small.

In graphene Dirac fermions have a high Fermi velocity,
$v_{F}=c/300$. Due to its high carrier mobility, graphene
based ballistic transistors operating at room temperature
have already been fabricated.\cite{dai-transistor} In addition to these unusual
electronic properties of graphene, the observation
of anomalous quantum Hall effect and the possibility of Klein
paradox are features, which attract the interest of researchers.
Electronic properties of graphene and graphene-based structures
have recently been reviewed.\cite{ geim-novo2007,revmod}

In the polar structures, such as BN, GaAs, after charge transfer
$p_{z}$ orbital electrons are located predominantly on one type of
atom. Thus the degeneracy of valence and conduction bands at $K$
point is removed and gap opening occurs.\cite{harrison} In Table 
\ref{table} the minimum width of band gaps calculated with LDA 
are given together with the symmetry points where the maximum 
(minimum) of valence (conduction) bands occur. Values of these 
band gaps after a correction by the GW$_{0}$ method are also 
given. The bands of compounds before and after GW$_{0}$ 
correction are also illustrated in Fig. \ref{fig6}.

Binary compounds have polar character in addition to the covalency
of bonds. Effective charge on cation and anion $Z_{c/a}^{*}$, 
charge transferred from cation to anion, $\delta \rho 
=Z_{a}^{*}-Z_{v}$ ($Z_{v}$ being the valency of the constituent 
atom) are calculated using Bader analysis. In spite of the 
ambiguities in finding the true effective charge, the calculated 
effective charges in Table \ref{table} give some idea about the 
direction of charge transfer and ionicity of the honeycomb 
structure. For some binary compounds like SiC, BN, AlN calculated 
effective charges appear to be right in sign but exaggerated in 
magnitude. We note that as the difference in the row numbers of 
constituent elements increases, $Z^{*}$ usually decreases. One 
can also generalize that the charge transfer decreases with 
increasing row number or atomic radii of anion if the cation is 
fixed. This trend is obvious in the structures of InN, InP, InAs and InSb.

\section{Heterostructures}

Depending on the constituent elements the band gaps of compound
honeycomb structures change in a wide energy range. In contrast,
the lattice constant $a$ of the compounds do not show significant
variation. The situation, where band gaps of two honeycomb
structures are significantly different, while their lattice
constants are practically the same, is a convenient condition to
make semiconductor heterostrucures. As an example, let us consider
AlN and GaN, which have LDA band gaps of 3.08 eV and 2.27 eV,
respectively. Their lattice constants are not significantly
different and are 3.09 and 3.20 \AA, respectively. Moreover, 
armchair nanoribbons can form pseudomorphic heterostructure with 
perfect junction. This is reminiscent of an AlN/GaN 
commensurate heterostructure having 2D interface. Owing to charge 
transfer between constituent nanoribbons at the junction, the 
bands are shifted and eventually aligned. Heterostructures of 
elemental and compound semiconductors generating a 2D electron 
gas and devices produced therefrom have been an active field of 
study in device physics in the past decades. It is expected that 
the heterostructure of armchair nanoribbons of GaN and AlN can 
constitute a 1D analog. When periodically repeated, this 
heterostructure can form superlattices behaving as multiple 
quantum wells or quantum dots. Earlier similar effects have been 
investigated for the heterostructures of graphene nanoribbons 
with different widths.\cite{graphene_applications4}

\begin{figure}
\includegraphics[width=8.5cm]{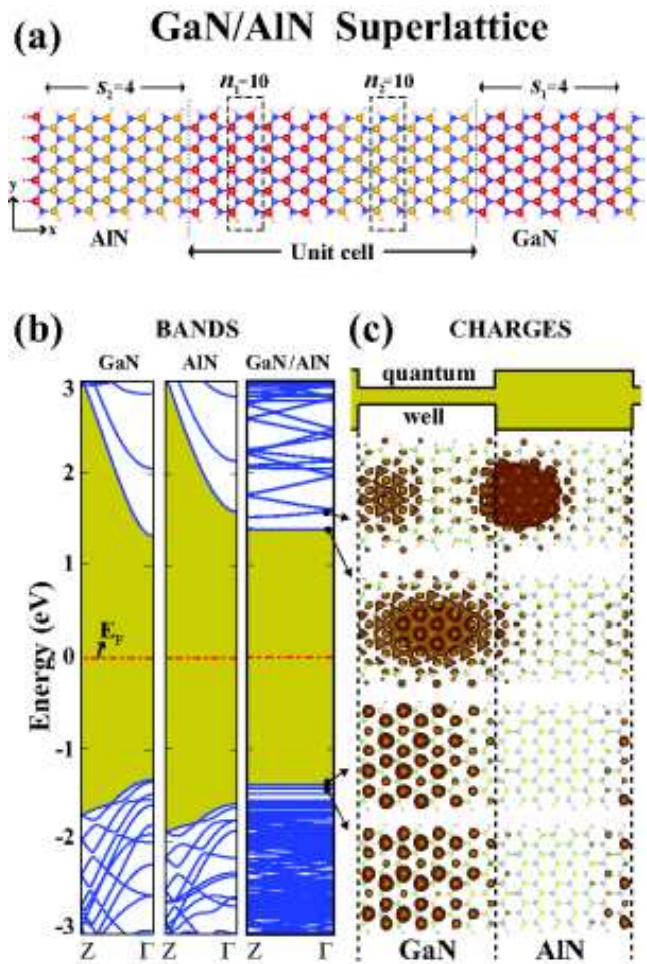}
\caption{(Color online) (a) A superlattice formed by periodically
repeating heterostructure of armchair nanoribbons of GaN-AlN compounds. Red, yellow, blue and small balls represent Ga, Al, N and H atoms, respectively. (b) Energy band structures of constituent GaN, AlN nanoribbons and resulting superlattice in momentum space. The band gaps are shaded by yellow. (c) Band decomposed isosurface
charge densities for lowest two conduction band and highest
two valence bands. Charges of lowest conduction band
and highest valence band states are confined in GaN side of the
junction, which has relatively smaller band gap than AlN.} \label{fig7}
\end{figure}

Superlattices of armchair honeycomb nanoribbon structures can be 
constructed according to the width and repeat periodicity of the 
constituent segments. We can label GaN/AlN superlattices as 
GaN/AlN($n_{1}$,$n_{2}$;$s_{1}$,$s_{2}$). Here, $s_{1}$ and 
$s_{2}$ specify the length of segments in terms of the numbers of 
the unitcells of constituent nanoribbons. Also $n_{1}$ and 
$n_{2}$ specify the width in terms of the number of dimer lines 
in the primitive unit cell of constituent nanoribbons. By varying 
the $n$ and $s$ , we can construct variety of superlattice 
structures. As a proof of concept, we consider a superlattice 
GaNAlN(10,10;4,4) as shown in Fig. ~\ref{fig7}. In the same figure 
we also presented the electronic band structure of constituent 
GaN and AlN nanoribbons. Upon construction, the atomic structure 
is fully optimized. Resulting energy band structure and charge 
density isosurfaces are presented in the same figure.

The highest valence band and the lowest conduction band 
states are flat and they are identified as \emph{confined} states.  As a result, one can deduce a type-I (normal) band alignment since states are confined to the GaN part of heterostructure. One notes that the bandgap of the superlattice in momentum space is different from those of constituent nanoribbons  and can also expect that the superlattice band gap in momentum space gets larger as the extension of GaN and AlN sides increases. Similar confined states can be obtained by constructing AlN core and GaN shell structures, where electrons are expected to be confined in the core region.

\section{Discussions and conclusions}

In view of the exceptional electronic, magnetic and mechanical properties of
recently synthesized graphene, questions have been raised whether
well-known materials in micro and optoelectronic industry can
attain similar honeycomb structures. It is hoped that
unusual properties can be attained from these structures. The
present paper examined a large number of materials, Group IV
elements, binary compounds of these elements, as well as a large
number of Group III-V  compounds to
reveal whether they may form 2D honeycomb structure. For several
decades the bulk crystals of these materials have dominated
micro and optoelectronic industry. Based on ab-initio structure
optimization and calculations of phonon modes we are able to
determine 22 honeycomb structure, which can be stable in a
local minimum on the Born-Oppenheimer surface as
either 2D infinite periodic crystals or finite size flakes
(patches). Our calculations reveal that Group IV elements, Si and
Ge, and binary compounds SiC, GeC, SnC, SnSi, SnGe and SiGe have
stable honeycomb structures. However, while SiC, GeC, SnC are
planar like graphene and BN, Si, Ge, SnSi, SnGe and SiGe are
buckled (or puckered) for stabilization. We also find that all
III-V compounds containing first row elements B, C or N
have planar stable structures. However, the binary compounds 
formed from the combination of {Al, Ga, In} and {P, As, Sb} are 
found to be stable in low buckled structure.

For honeycomb structures which were deduced to stable, an 
extensive analysis have been carried out to determine their 
atomic structure, elastic and electronic properties. While Si and 
Ge are semimetallic and linear band crossing at the Fermi level 
like graphene, all the binary compounds are found to be 
semiconductors. Interestingly, these honeycomb materials exhibit 
interesting trends regarding cohesive energy, band gap, effective 
charge, in-plane stiffness, Poisson's ratio depending on the row 
numbers of their constituent elements or their radius. 
Interestingly, we deduced a relation between in-plane stiffness 
and cohesive energy among all honeycomb structures studied in 
this work.

These materials in honeycomb structure have a variety of
band gaps. Even more remarkable is that the nanoribbon forms of 
these materials provide diverse properties depending on not only 
their constituents, but also their chirality and width. All these 
properties are expected to offer number of applications. 
Therefore, the studies related with their functionalization  by 
vacancy defects or adatoms, their mechanical and spintronic 
properties, their heterostructures and core-shell structures
will open a new field of research. We hope that the findings in this work will promote the research aiming at the synthesis of these materials.

\begin{acknowledgments}

Computing resources used in this work were provided by the
National Center for High Performance Computing of Turkey (UYBHM)
under grant number 2-024-2007. This work was partially supported by T\"{U}BITAK under grant no: 106T597. S. C and R. T. S acknowledge
financial support from T\"{U}BA (Turkish Academy of Sciences).
\end{acknowledgments}

\end{document}